\def\half{\frac{1}{2}}
\def\hi#1#2{$#1$\kern -2pt-#2} 
\def\hy#1#2{#1-\kern -2pt$#2$} 
\def\dbox#1{\hbox{\vrule 
\vbox{\hrule \vskip #1\hbox{\hskip #1\vbox{\hsize=#1}\hskip #1}\vskip #1 
\hrule}\vrule}}  
\def\qed{\begin{flushright}~{\dbox{0.05true in}}\end{flushright}} 

\documentclass[showkeys,showpacs]{revtex4}
\usepackage{graphicx}
\begin{document}
\hspace*{5 in}CUQM-134

\title{Relativistic comparison theorems}
\author{Richard L. Hall}
\address{Department of Mathematics and Statistics, Concordia University,
1455 de Maisonneuve Boulevard West, Montreal,
Quebec, Canada H3G 1M8}
\email{rhall@mathstat.concordia.ca}
\begin{abstract}Comparison theorems are established for the Dirac and Klein--Gordon equations. 
We suppose that $V^{(1)}(r)$ and $V^{(2)}(r)$ are two real attractive central potentials in $d$ dimensions
that support discrete Dirac eigenvalues $E^{(1)}_{k_d\nu}$ and $E^{(2)}_{k_d\nu}$.  We prove that if $V^{(1)}(r) \leq V^{(2)}(r)$, then each of the corresponding discrete eigenvalue pairs  is ordered $E^{(1)}_{k_d\nu} \leq E^{(2)}_{k_d\nu}$. This result generalizes an earlier more restrictive theorem that 
required the wave functions to be node free. For the  the Klein--Gordon equation, similar reasoning also leads to a comparison theorem provided in this case that the potentials are negative and the eigenvalues are positive.
\end{abstract}
\vskip0.2in
\maketitle
\section{Introduction}
The comparison theorem of quantum mechanics states that if two real potentials are ordered, $V^{(1)}(r) \leq V^{(2)}(r),$ then each
corresponding pair of eigenvalues is ordered $E^{(1)}_{\ell n} \leq E^{(2)}_{\ell n}.$  For non-relativistic problems whose Hamiltonians are bounded below, this theorem is an 
immediate consequence of the variational characterization of the discrete spectrum. By contrast, since the energy operators for relativistic problems are not bounded below, a simple comparison theorem is  unexpected since the usual variational arguments are not applicable \cite{franklin,goldman,grant}. However, by  using reasoning developed originally for Schr\"odinger problems in which the graphs of the comparison potentials cross each other in a controlled way \cite{hallrcom}, and consequently variational inequalities are no longer available, it was possible to derive a limited comparison theorem \cite{halld} valid for node-free Dirac eigenstates; this was subsequently extended to Dirac problems  in $d$ dimension \cite{chen}. In a more recent development \cite{hallds}, it was shown for the Dirac problem that if an attractive potential $V(r,a)$ is monotone in a potential parameter $a$, then each corresponding Dirac eigenvalue $E(a)$ is monotone in this parameter. Both types of result have also been proved for potentials in the Klein--Gordon equation \cite{hallkg}, with the added restriction in this case that the eigenvalues considered are positive.  We shall see in this paper that the monotonic spectral behavior induced by attractive central potentials $V(r,a)$ that are monotonic in a parameter $a$ leads to a general comparison theorem for the Dirac problem, and to a comparison theorem restricted to negative potentials and positive eigenvalues for the Klein--Gordon equation.
These results are established below in sections~II and III respectively.  In section~IV we consider an example of comparison approximations to the Dirac spectrum of a screened-Coulomb potential: in particular this establishes a general energy-bound conjecture  posed 25 years ago in Ref.~\cite{halldsc} and later proved to be true for node-free states in Ref.~\cite{halld}.

The potential $V(r)$ discussed in this paper is sometimes called a `vector potential' since it enters the problem as the time component of a $4$-vector: it is this potential that for Hydrogen-like atomic models is taken to be $V(r) = -Z\alpha/r.$  A `scalar potential' $S(r)$ may also be considered, as a variable term added to the mass \cite{greiner}.  However, throughout the present paper, for both Dirac and Klein--Gordon problems, we shall assume that $S = 0$ and that the mass $m$ is constant.

\section{The Dirac equation} 
For a central potential $V(r)$ in $d$ dimensions the Dirac equation can be written \cite{jiang} in natural units $\hbar=c=1$ as
\begin{equation}\label{eq3}
i{{\partial \Psi}\over{\partial t}} =H\Psi,\quad {\rm where}\quad  H=\sum_{s=1}^{d}{\alpha_{s}p_{s}} + m\beta+V,
\end{equation}
where $m$ is the mass of the particle, and $\{\alpha_{s}\}$ and $\beta$  are Dirac matrices, which satisfy anti-commutation relations; the identity matrix is implied after the potential $V$. For stationary states, algebraic calculations in a suitable basis \cite{jiang} lead to a pair of first-order linear differential equations in two radial functions $\{\psi_1(r), \psi_2(r)\}$, where $r = |\mathbf{r}|.$  For $d > 1,$ these functions vanish at $r = 0$, and, for bound states, they may be normalized by the relation 
\begin{equation}\label{eq4}
(\psi_1,\psi_1) + (\psi_2,\psi_2) = \int\limits_0^{\infty}(\psi_1^2(r) + \psi_2^2(r))dr = 1.
\end{equation}
We use inner products {\it without} the radial measure factor $r^{(d-1)}$ because the factor $r^{\frac{(d-1)}{2}}$ is already built in to each radial function. Thus the radial functions vanish at $r = 0$ and satisfy the coupled equations
\begin{eqnarray}
E\psi_1 &=& (V+m)\psi_1 + (-\partial + k_{d}/r)\psi_2\label{eq5}\\
E\psi_2 &=& (\partial + k_{d}/r)\psi_1 + (V-m)\psi_2\label{eq6},
\end{eqnarray}
where $k_1 = 0,$ $k_{d}=\tau(j+{{d-2}\over{2}}),~d >1$, $\tau = \pm 1,$  and $E = E_{k_d\nu}$ is the eigenvalue corresponding to the state with $\nu = 0,1,2,\dots$ nodes in the upper radial function $\psi_1(r).$   We note that the variable $\tau$ is sometimes written $\omega$, as, for example in the book by Messiah \cite{messiah}, and the radial functions are often written $\psi_1 = G$ and $\psi_2 = F,$ as in the book by Greiner \cite{greiner}.  We shall assume that for each attractive central potential $V$considered, there is a discrete eigenvalue $E = E_{k_d\nu}$ and that Eqs.(\ref{eq5},~\ref{eq6}) are the eigenequations for the corresponding radial eigenstates.  In this paper we shall present the problem explicitly for the cases $d > 1.$ Similar arguments go through for the case $d=1$: in this case $k_1 = 0,$ the states can be classified as even or odd, and the normalization (\ref{eq4}) becomes instead $\int_{-\infty}^{\infty}\left(\psi_1^2(x) + \psi_2^2(x)\right)dx = 1.$ 
 We first suppose that the potential $V=V(r,a)$ depends smoothly on a parameter $a$, and we re-state a theorem from Ref.~\cite{hallds}:

\medskip
\noindent{\bf Theorem~1}~~{\it The real attractive central potential $V(r,a)$ depends smoothly on the parameter $a$, and $E(a)= E_{k_d\nu}(a)$ is a corresponding discrete Dirac eigenvalue. Then:}
\begin{equation}\label{eq1}
\partial V/\partial a \geq 0~~\Rightarrow~~E'(a) \geq 0~~~{\it and}~~~\partial V/\partial a \leq 0~~\Rightarrow~~E'(a) \leq 0.
\end{equation} 
The principle result of this section of the present paper is the proof of the following general comparison theorem.

\medskip
\noindent{\bf Theorem~2}~~{\it Suppose that $E^{(1)}_{k_d\nu}$ and $E^{(2)}_{k_d\nu}$ are Dirac eigenvalues corresponding to two distinct 
attractive central potentials $V^{(1)}(r)$ and $V^{(2)}(r)$.  Then: }
\begin{equation}\label{eq2}
V^{(1)}(r) \leq V^{(2)}(r) ~~~ \Rightarrow ~~~E^{(1)}_{k_d\nu} \leq E^{(2)}_{k_d\nu}.
\end{equation}

\medskip
\noindent{\bf Proof of Theorem~2:}~~The theorem may be proved by keeping the given real comparison potentials $V^{(1)}(r)$ and $V^{(2)}(r)$ fixed and defining the one-parameter family 
of potentials $V(r,a)$ by
\begin{equation}\label{eq7}
V(r,a) = V^{(1)}(r) + a\left(V^{(2)}(r) - V^{(1)}(r)\right),\quad a \in[0,1].
\end{equation}
We now consider one of the discrete eigenvalues, $E_{k_d\nu}(a) = E(a),$ of the Dirac equation with potential $V(r,a).$ In terms of this energy function, the comparison eigenvalues of the theorem are given by $E^{(1)}_{k_d\nu} = E(0)$ and $E^{(2)}_{k_d\nu} = E(1).$  Clearly, since $V^{(2)}(r) - V^{(1)}(r) \ge 0,$ we have $\partial V(r,a)/\partial a = V^{(2)}(r) - V^{(1)}(r) \ge 0.$  Thus, by Theorem~1 we have $E'(a) \ge 0.$ The fact that $E(a)$ is increasing in turn implies that $E(0) \le E(1),$ which result establishes the theorem. \qed 

\section{The Klein--Gordon equation}
The Klein--Gordon equation for an attractive central potential $V(r,a)$ depending on a parameter $a$ is given by
\begin{equation}\label{eqkg}
(-\Delta +  m^2)\psi = (E -V)^2\psi.
\end{equation}
The generalization to arbitrary dimension $d >1$ is conveniently allowed for \cite{hallkg} by replacing the orbital-angular momentum quantum number $\ell$ of the three-dimensional problem by $\ell_d$, where
\[
\ell_d = \ell + (d-3)/2.
\]
The eigenvalues of the problem in $d$ dimensions are labelled by $\ell_d$ and the number $\nu = 0,1,2,\dots$ of nodes in the radial eigenfunction $\psi(r).$ The  radial equation becomes from (\ref{eqkg})
\begin{equation}\label{eqkgr}
-\psi''(r) + \frac{Q}{r^2}\psi(r) = \left(\left(E_{\ell_d\nu}-V(r)\right)^2 -m^2\right)\psi(r), 
\end{equation}
where
\begin{equation}\label{eqkgq}
Q=\frac{1}{4}(2\ell+d-1)(2\ell+d-3) = \ell_d(\ell_d+1),\quad \ell = 0,1,2,\dots,\ d = 2,3,\dots
\end{equation} 
In Ref.\cite{hallkg} the following theorem was proved:

\medskip
\noindent{\bf Theorem~3}~~{\it The real attractive central potential $V(r,a) \le 0$ depends smoothly on a parameter 
$a$, and $E(a)= E_{\ell_d\nu}(a)\ge 0$ is a corresponding discrete Klein-Gordon eigenvalue. Then:}
\begin{equation}\label{eqth3}
\partial V/\partial a \ge 0 \Rightarrow E'(a) \ge  0,\quad {\it and}\quad \partial V/\partial a \le 0 \Rightarrow E'(a) \le  0.
\end{equation}

\medskip
\noindent  By similar reasoning to that used for the Dirac case, we shall now prove the Klein--Gordon comparison theorem:

\medskip
\noindent{\bf Theorem~4}~~{\it Suppose that $E^{(1)}_{\ell_d\nu}$ and $E^{(2)}_{\ell_d\nu}$ are Klein--Gordon eigenvalues corresponding to two distinct 
attractive central potentials $V^{(1)}(r)$ and $V^{(2)}(r)$.  Then: }
\begin{equation}\label{eqth4}
V^{(1)}(r) \leq V^{(2)}(r) \le 0~~{\it and}~~E^{(1)}_{\ell_d\nu} > 0~~~ \Rightarrow ~~~E^{(1)}_{\ell_d\nu} \leq E^{(2)}_{\ell_d\nu}.
\end{equation}

\medskip
\noindent{\bf Proof of Theorem~4:}~~We suppose that $V^{(1)}(r) \leq V^{(2)}(r) \le 0$ and we define a one-parameter family of negative potentials $V(r,a)$ by 
\begin{equation}\label{eqkgpot}
V(r,a) = V^{(1)}(r) +  a\left(V^{(2)}(r) - V^{(1)}(r)\right),\quad a \in[0,1].
\end{equation}
Clearly, $V(r,a) \leq 0$.  Meanwhile, $\partial V(r,a)/\partial a = V^{(2)}(r) - V^{(1)}(r)\ge 0.$ We also suppose that $E^{(1)}_{\ell_d\nu} = E(0)$ is positive.
By Theorem~3 we deduce that $E'(a) \ge 0$ for $a \in [0,1].$  Thus we conclude that
$$E^{(2)}_{\ell_d\nu} = E(1) \ge E(0) = E^{(1)}_{\ell_d\nu},$$
 which inequality establishes the theorem.\qed

\section{Energy  bounds for a screened--Coulomb potential}
We revisit an example discussed in Refs.~\cite{halld,halldsc},  but now we consider the problem in general spatial dimension $d >1$, and we are able to construct energy bounds valid for every discrete eigenvalue. We first consider the Dirac equation for the pure Coulomb problem with potential $V(r) = -u/r,$ where the coupling parameter $u=\alpha Z$ is not too large. We write the exact discrete eigenvalues as $D_{k_d\nu}(u) = D(u)$ and they are given \cite{jiang,dong} exactly by 
\begin{equation}\label{eqdirach}
D(u) = \left\{1 + u^{2}\left[\nu + (k_d^{2} - u ^{2})^{\half}\right]^{-2}\right\}^{-\half},
\end{equation}
where 
\begin{equation}\label{eqtau}
k_d^2 = \left(j + \frac{d-2}{2}\right)^2
\end{equation}
and $\nu = 0,1,2, \dots$ counts the nodes in the upper radial function $\psi_1(r)$. The principal quantum number $n$ is defined in general as 
\begin{equation}\label{eqpqn}
n =  \nu +|k_d| - \frac{d-3}{2}
\end{equation}
and the spectroscopic designation 
\[
\{s,p,d,\dots\}\leftrightarrow \ell = \{0,1,2,\dots\}
\]
 is provided by the formula
\begin{equation}\label{ell}
\ell = |k_d| - \left(\frac{d-1}{2}\right).
\end{equation}
With these conventions, the eigenvalue formula (\ref{eqdirach}) becomes formally similar to the expression for the
 three-dimensional case:
\begin{equation}\label{eqDu}
D(u) = \left\{1 + u^{2}\left[n-\ell-1 + (k_d^{2} - u ^{2})^{\half}\right]^{-2}\right\}^{-\half}.
\end{equation}

The comparison theorem established in this paper allows  us to  use the exact eigenvalues of the Coulomb problem, for example, to approximate those of a related problem that is nearby in a different sense from what is normally used in perturbation theory.  Specifically, we consider the Mehta--Patil screened-Coulomb potential $V(r)$ given \cite{mp78} by
\begin{equation}\label{eqmppot}
V(r) = -\left(\frac{v}{r}\right)\left[1-(1-1/Z)\frac{\lambda r}{1+\lambda r}\right],
\end{equation}
where, for example, the appropriate potential parameters for an atomic model in three dimensions are
\begin{equation}\label{eqmpparams}
v = \alpha Z\quad{\rm and}\quad \lambda =0.98\,\alpha Z^{\frac{1}{3}}.
\end{equation}
The comparison theorem may  now be invoked because we can show that, for each value of the parameter $t >0$, $V(r)$ is bounded above by a shifted-Coulomb potential $V^{(t)}(r)$ of the form 
\begin{equation}\label{eqVt}
V(r) \le V^{(t)}(r) = -\frac{a(t)}{r} + b(t).
\end{equation}
This follows by the following argument from envelope theory \cite{hallenv,hallsc,hallpow}. If we write $V(r)$ as a transformation $V(r) = g(h(r))$ of the hydrogenic potential $h(r) = -1/r,$ then the transformation function $g(h)$ becomes 
\begin{equation}\label{eqgh}
g(h) = v\left[h + \lambda(1-1/Z)\left(1+\frac{\lambda}{h-\lambda}\right)\right].
\end{equation}
It follows immediately that the function $g(h)$ is monotone, $g'(h) > 0$, and concave, $g''(h) < 0.$  This means that $g(h)$ lies above its tangents, all of which are of the form $a(t)h(r) + b(t)$, with $a(t)>0,$  that is to say, shifted attractive Coulomb potentials. The coefficients are given by
\begin{equation}\label{eqabt}
a(t) = g'(h(t)),\quad b(t) = g(h(t)) - h(t)g'(h(t)),
\end{equation}
where $r = t$ is the point of contact between $V^{(t)}(r)$ and $V(r).$  For each of these upper tangential potentials, the spectrum is given exactly (with the aid of Eq.~(\ref{eqDu})) by the right-hand side of the equation
\begin{equation}\label{eqEt}
E \le D(a(t)) + b(t).
\end{equation}
This inequality is a consequence of the potential inequality $V(r) \le V^{(t)}(r)$ and Theorem~2. 
If we optimize over $t >0,$ and effect the change of variable $t\rightarrow u = g'(h(t))$, then the critical point is found to be the same as that of the energy function ${\mathcal E}(u)$ given by

\[
{\mathcal E}(u) = D(u)- uD'(u) + V(-1/D'(u)).
\]

\noindent Thus we obtain the following 
best upper-bound formula, valid for each discrete Dirac eigenvalue, and expressed in terms of the pure hydrogenic spectral function $D(u)= D_{k_d\nu}(u)$ of Eq.~(\ref{eqDu}) by

\begin{equation}\label{eqEu}
E \le \min_{u > 0}\left[D(u)- uD'(u) + V(-1/D'(u))\right].
\end{equation}
This same formula yields an upper energy  bound for {\it any} potential $V(r)$ that is a  monotone increasing and concave function of $h(r) = -1/r$. For potentials that are {\it convex} functions of $h(r)$, the same expression provides a {\it lower} energy bound.  Some numerical values for node-free states in dimension $d=3$ are given  in Ref.~\cite{halld}, and a more extensive table of results is given in Ref.~\cite{halldsc}: at that time, the data in the latter table could only be considered as {\it ad hoc} approximations, although the inequality  Eq.~(\ref{eqEu}) was proposed then as a conjecture.  Since it follows from Theorem~2 that Eq.~(\ref{eqEu}) is valid for all the discrete eigenvalues and for arbitrary spatial dimension $d > 1$, we may now state, in particular, that the values given in Table~1 of Ref.~\cite{halldsc} are {\it a priori} all upper bounds.  We make weaker general claims for the Klein--Gordon equation: for the screened-Coulomb example (where $g$ is concave), $D(u)$ would be set to the exact Klein--Gordon Coulomb energy, and the energy bound (\ref{eqEu}) is then known, by Theorem~4, to be valid provided that the potential is negative and the eigenvalues considered are positive.

\section{Conclusion}
If we examine exact relativistic eigenvalue formulas given, for example, in the book by Greiner \cite{greiner}, 
we see that, whenever a potential depends monotonically on a parameter, the eigenvalues are monotonic in this parameter in the same direction.
Since the corresponding non-relativistic problems usually behave this way, we are not surprised.  However, the question immediately arises as to how general such
 spectral features are in relativistic quantum  mechanics.  In this paper, comparison theorems are proved for Dirac and Klein--Gordon problems in which a single particle is bound by a central potential.  These results are established without the use of any variational arguments. The Klein--Gordon result which we have been able to obtain is limited to negative potentials and positive eigenvalues.
In the Dirac case, there is no limitation save that the comparison potentials each yield discrete eigenvalues in the same angular-momentum sector and with the same number of radial nodes.  
The illustration, involving shifted-Coulomb potential that are upper bounds to a screened-Coulomb potential, 
 shows how the existence of a comparison theorem immediately leads to spectral approximations.  For many relativistic problems, this allows the type of spectral 
reasoning and estimates that are commonly employed in non-relativistic quantum mechanics.

 \section*{Acknowledgment}
Partial financial support of his research under Grant No.~GP3438 from the Natural
Sciences and Engineering Research Council of Canada is gratefully acknowledged. 

\end{document}